\documentclass{article}
\usepackage[utf8]{inputenc}
\usepackage{natbib}
\usepackage{color}
\usepackage{graphicx}
\usepackage{hyperref}
\usepackage{url}
\usepackage[section]{placeins}
\providecommand{\keywords}[1]
{
  \small	
  \textbf{\textit{Keywords---}} #1
}
\usepackage{geometry}
\title{Flight restrictions from China 
during the COVID-2019 Coronavirus outbreak}
\author{Stefano Maria Iacus$^*$ \and Fabrizio Natale$^*$ \and Michele Vespe$^*$ }
 
 \date{%
   \small{ $^*$European Commission, Joint Research Centre}\\[2ex]%
    \today
}

\begin{document}

\maketitle

\begin{abstract}
This short note provides estimates of the number of passengers that travel from China to all world airports in the period October 2019 - March 2020 on the basis of historical data. From this baseline we subtract the expected reduction in the number of passengers taking into account the temporary ban of some routes which was put in place since 23 January 2020 following the COVID-2019 Coronavirus outbreak. The results indicate a reduction of the number of passengers in the period January - March 2020 of -2.5\%. 
This calculation considers only the complete closure of routes (not just direct flights) and not the reduction in the number of passengers on still active direct and indirect connections. At the moment of writing, with such partial information it is premature to quantify economic losses on the civil air transport and tourism industry. This note is meant to provide a baseline that be extended to all countries of origin and updated as more recent data will become available.  

\section*{Highlights}
\begin{itemize}
   \item Independently from the Coronavirus outbreak, the growth of air passengers flows from China is  stabilising since 2018. 
    \item Almost 83\% of the flows are within China. 
   \item The COVID-2019 Coronavirus outbreak coincides with the period of the year characterised by the lowest level of air traffic from China. 
   \item Under normal circumstances the flows to the EU and UK were increasing from April to July with the approaching of the peak season for tourism.
    \item For the period January - March 2020 the reduction in flows can be estimated in -2.5\%. This estimate is only considering the complete closure of routes.
        \item From the data currently available it is premature to evaluate the economic impacts on the air transport network and tourism industry. Past analyses for SARS show that impacts linked to changes in consumer behaviour and dictated by fear can be considerably larger than the direct costs for the management of the emergency.
\end{itemize}
\end{abstract}

\keywords{COVID-2919, coronavirus, air passengers data}

\pagebreak
\section{Introduction}
The present note provides a preliminary analysis of the changes in the international air flows from China following the COVID-2019 Coronavirus outbreak.

At the time of writing the outbreak has caused 80,348 cases of infections and 2,707 deaths globally. Most of the cases are registered in China (96.6\%), followed by South Korea (1.2\%), the cruise ship Diamond Princess (0.86\%) and Italy (0.36\%) \citep{Worldometer}.
More than 70 countries have closed their borders or imposed entry bans on residents of China or foreign nationals who have visited China. Most airlines based in these countries have suspended or reduced their flights \citep{scmp1}.

The disruption of air traffic is expected to play a major role in a reduction of GDP growth at global level. At the G20 summit on 22 February 2020 in Riyadh, finance ministers and central bank governors have indicated in the ongoing Coronavirus epidemic a threat, which is putting global economic recovery at risk \citep{imf2020}. Some delegations see in this new epidemic a sign of fragility of economies which are increasingly exposed to inter-dependencies of supply chains and large international mobility for labour and tourism\footnote{The contribution of tourism is estimated in 2018 at 10.4\% of global GDP \citep{WTTC}}.

The number of cases of the current Coronavirus outbreak are already 10 times larger in respect of the 2013 SARS epidemic. In the case of SARS, economic losses were estimated at \$40 billion \citep{lee_impact_2003}. More than to the disease itself, these losses have been attributed to changes of behavior and fear, which propagate from the affected countries to the entire world economy. Losses entail short-term shocks on industries such as tourism, air transport, disruption of supply chains and longer term indirect effects on decline in consumer demand and country risk premium.  
For the air industry, IATA estimated a loss due to the SARS epidemic of \$6 billion revenue (8\% of the annual traffic) for the Asia-Pacific airlines and \$1 billion (3.7\% of annual total traffic) for North American airlines in a period of 8 months between November 2013 and June 2014 \citep{IATA2016}.

This note is not intended to give an economic assessment of impacts but only a quick and preliminary analysis while the epidemic is still progressing.

In particular the note provides statistical estimates:
\begin{itemize}
   \item of the flows of passengers from China for the period October - March 2020, considering the normal prolongation of historical trends;
   \item of the reduction linked to complete closure of routes since 23 January 2020, caused by the Coronavirus outbreak.
\end{itemize}

The reduction does not take into account of the lower number of passengers on still active direct and indirect connections, therefore it should be considered as an estimate from below of the true passengers volume reduction.

\section{Data}

The analysis makes use of the Air Passenger Dataset provided by SABRE, a private company that collects information on global air traffic \citep{sabre}.
The data contain consolidated monthly air traffic information on the number of passengers and the average price per ticket, for every couple of airports around the world for which a direct or indirect connection exists, i.e., the data register the actual flight volumes from each true origin and true final destination, taking into account intermediate stops. 
The most recent SABRE data  comprise the period January 2010 until September 2019.
The data considered in our analysis consist of all international flights originating from China (mainland), toward all countries in the world and all domestic connections within China. The number of distinct airports dyads amounts to 25,681 connections, from 282 origin airports and 1,622 true destination airports.

\section{Methods}

To anticipate the release of more recent data from SABRE, we forecast the air volumes for the period October 2019 - March 2020 on the basis of historical trends between 2010 and September 2019. For the forecast we use a non-homogeneous Poisson process with a periodic intensity  function calibrated on historical data \citep{IacusYoshida2018} and extrapolated non-linearly for the future years, e.g., the time series of all January's from 2010 till 2019 is used to calibrate a regression model with quadratic terms and the value for 2020 is projected accordingly. The intensity function of the non-homogeneous Poisson process is therefore the combination over all periods. This approach is needed to  take into account that air passengers volumes show non-linear trends and cycles  which are specific to each route, so a model has been fitted for each route as well.

With this method we were able to forecast both air flows and overall ticketing revenues\footnote{Although, we do not discuss the economic impact on aviation in this paper.} as if no flight/route were stopped. In fact, as of 23 January 2020, a number of (direct) routes from Wuhan have been closed, but not only those flights. To have an estimate of the number of passengers lost due to flight suppression, we use a web-scraping strategy to look for all flights available corresponding to the 25,681 unique dyads in our data set. We used Kiwi\footnote{\texttt{Skypicker} are the API server of \texttt{Kiwi.com}}, which is a service used by travel agencies and individuals for on-line flight booking. It turns out that around 1,000 unique connections from China have zero flights in the period 7-25 February 2020 and thus we take it as a proxy of the real number of routes temporary cancelled due to COVID-2019 outbreak.

\section{Baseline estimates of the air passengers flows until March 2020}
\begin{figure}[!htb]
\centering{\includegraphics[width=1\textwidth,height=0.4\textheight]{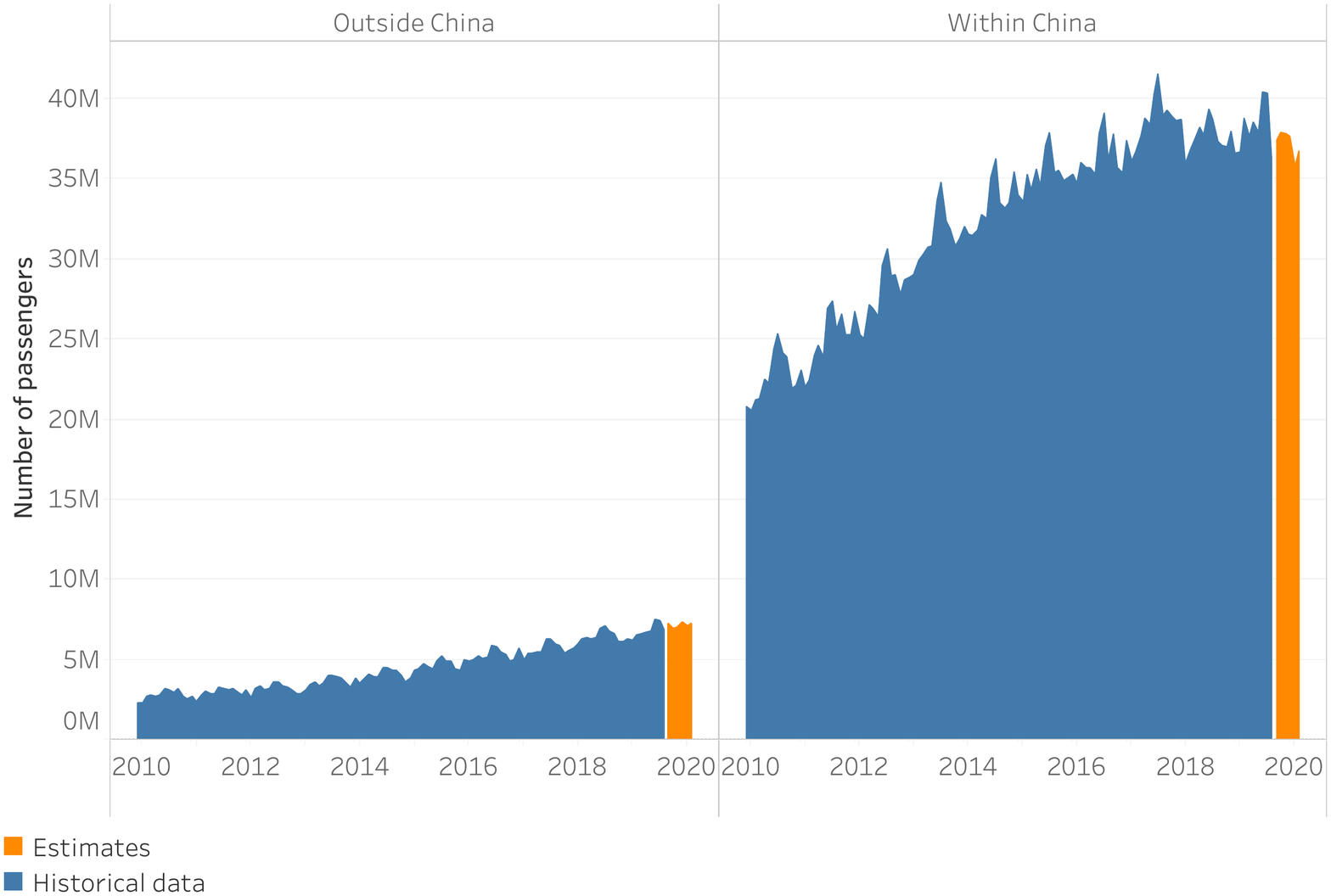}}
\caption{Monthly number of passengers from Chinese airports within and outside China. Source: SABRE data set (until September 2019) and own estimates based on historical trends (October 2019 - March 2020).}
\label{fig1}
\end{figure}

The trends for the number of air passengers from Chinese airports are constantly increasing since 2010 (Figure~\ref{fig1}). 
The volume of monthly traffic reached a peak of 46 million passengers in July 2017. After this peak the traffic in particular within China is stabilising.

Following this trend our estimates indicate that the number of monthly passengers in the period January - March 2020, in the absence of Coronavirus, would have been of around 43.9 million on average.
Almost 83\% of the flows are directed to airports within China. 

\begin{figure}[!htb]
\centering{\includegraphics[width=1\textwidth,height=0.4\textheight]{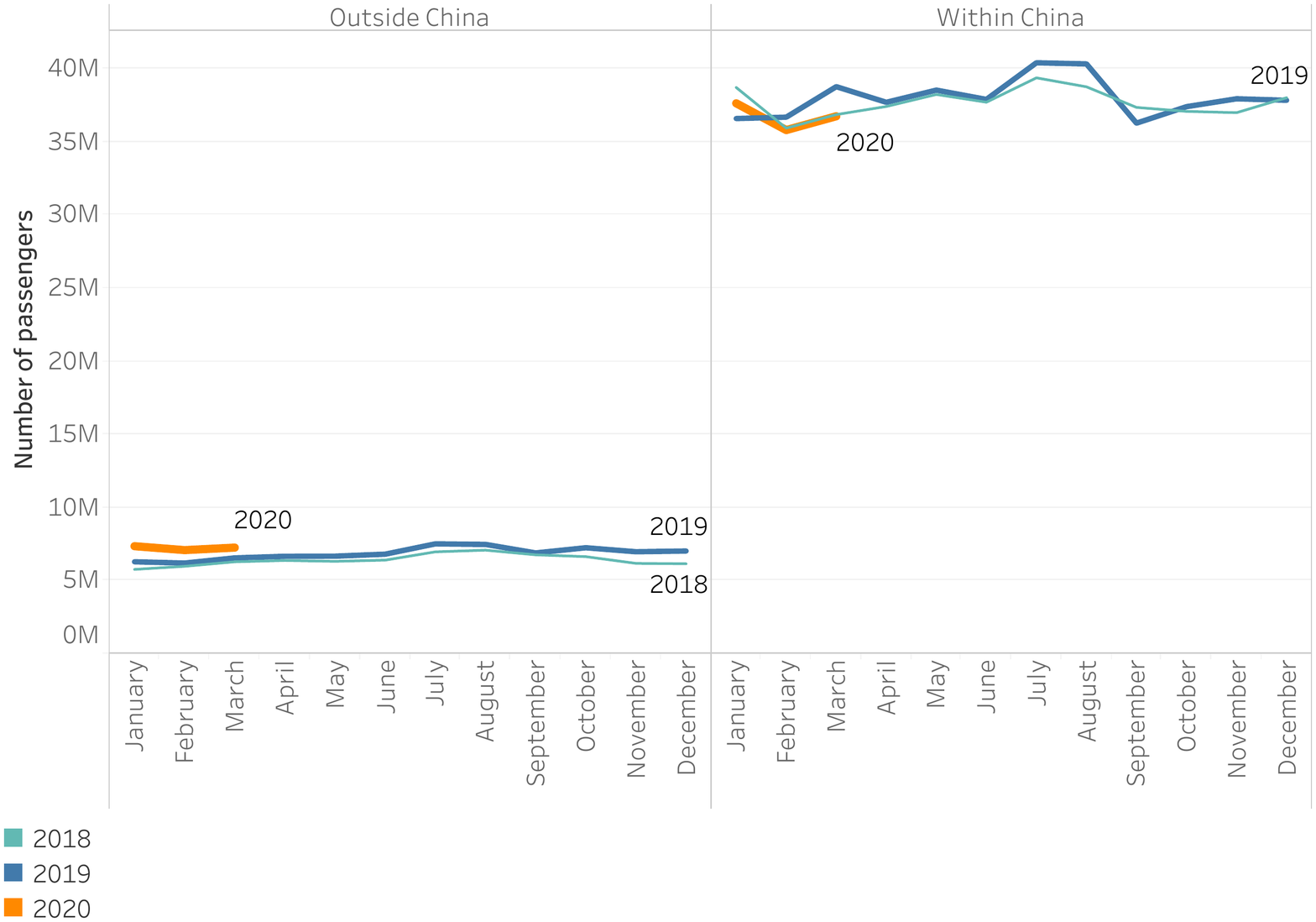}}
\caption{Monthly number of passengers from Chinese airports within and outside China in 2018, 2019 and 2020. Source: SABRE data set (until September 2019) and own estimates based on historical trends (October 2019 - March 2020).}
\label{fig2}
\end{figure}

In the period January - March 2020, in the absence of effects from the Coronavirus outbreak, we estimate an increase in the number of passengers for flights outside China with similar monthly distribution to previous years (Figure~\ref{fig2}). On the contrary, for flights within China our baseline model is  showing a reduction to the levels of 2018. This reduction can be attributed to a slow down of the economic growth in China, which was anticipated by IMF independently from the Coronavirus outbreak \citep{IMF}.

\begin{figure}[!htb]
\centering{\includegraphics[width=1\textwidth,height=0.4\textheight]{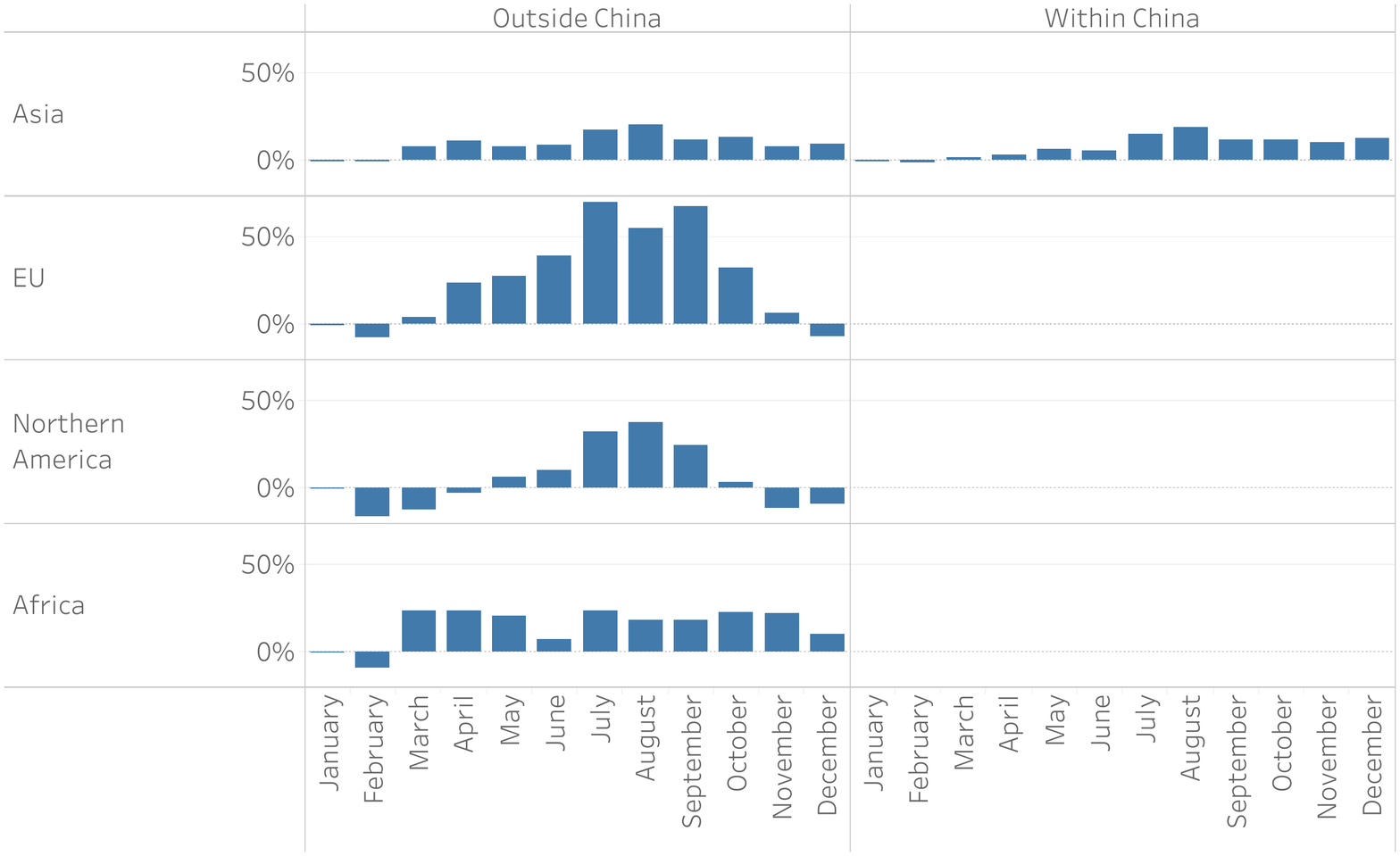}}
\caption{Differences in the volume of passengers in respect of the reference month of January by region of destination. The differences are calculated considering data for the entire period 2010-2019.}
\label{fig3}
\end{figure}

The flows to Asia, Africa and within China do not exhibit strong variations across months (Figure~\ref{fig3}). A seasonality in the monthly distribution of flows is more evident in the case of flights towards the EU and North America. On average February is the month with the lowest level of traffic (-16\% compared to January in the case of Northern America and -7.7\% in the case of EU). Flights towards the EU, UK and Northern America gradually increase during spring and reach e peak during the summer months. As introduced by \cite{gabrielli2019dissecting}, the seasonal component of monthly air passenger traffic can be linked to different types of mobility, from tourism to seasonal work migration. In this case, such differences in seasonality may be explained by the prevalence of business and labor mobility from/to Asia and Africa, and tourist flows from/to the EU and North America.  

\begin{figure}[!htb]
\centering{\includegraphics[width=1\textwidth,height=0.4\textheight]{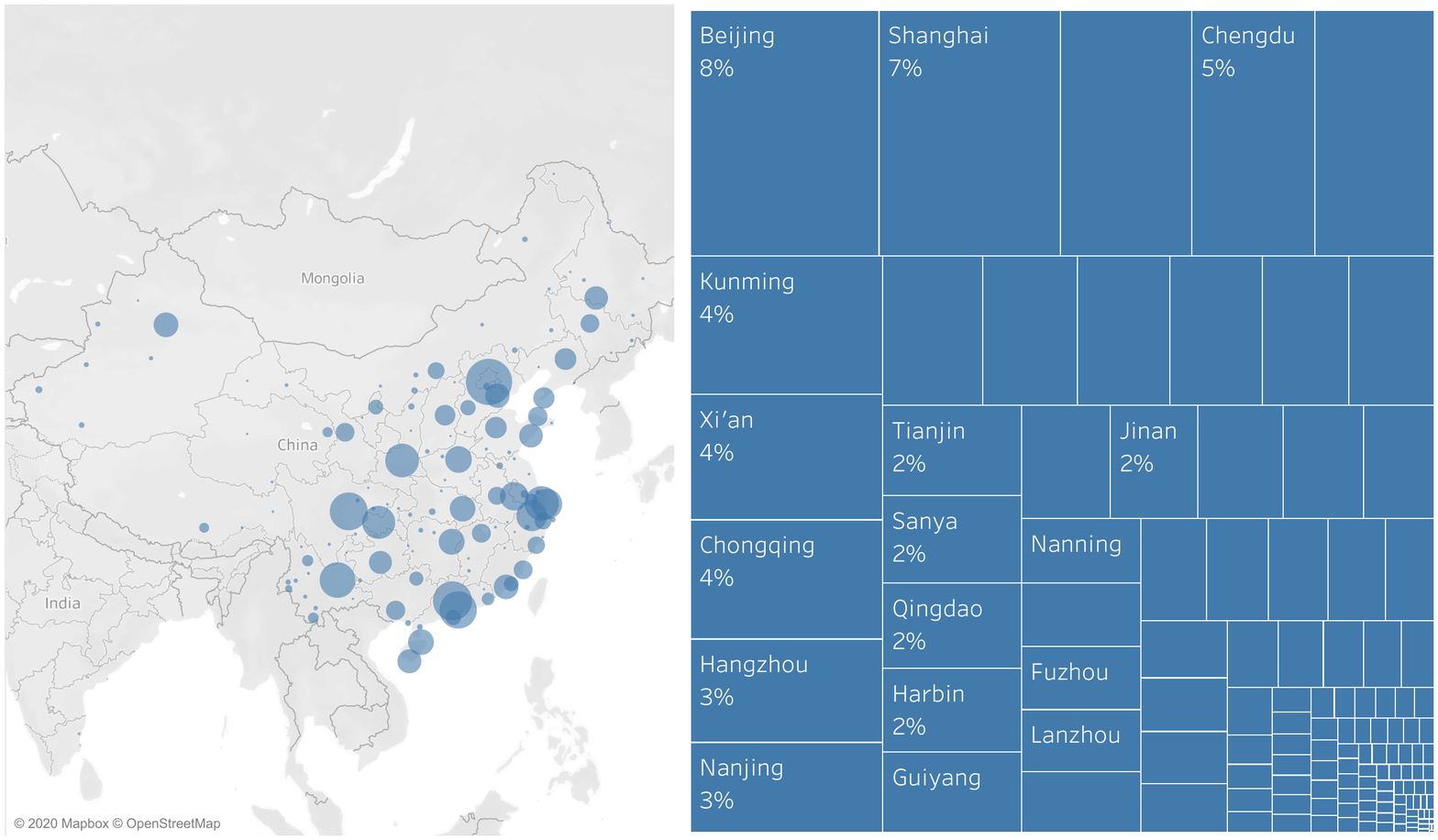}}
\caption{Main destinations within China for flights originating from China in the period January - April 2020 (baseline without effects from the Coronavirus epidemic).}
\label{fig4}
\end{figure}

In terms of geographical distribution the baseline estimates for the first three months of 2020 indicate that flows within China are mostly directed to  the airports of Beijing (7.3\%), Shanghai (7.1\%) and Guangzhou (5.2\%) (Figure~\ref{fig4}).

\begin{figure}[!htb]
\centering{\includegraphics[width=1\textwidth,height=0.4\textheight]{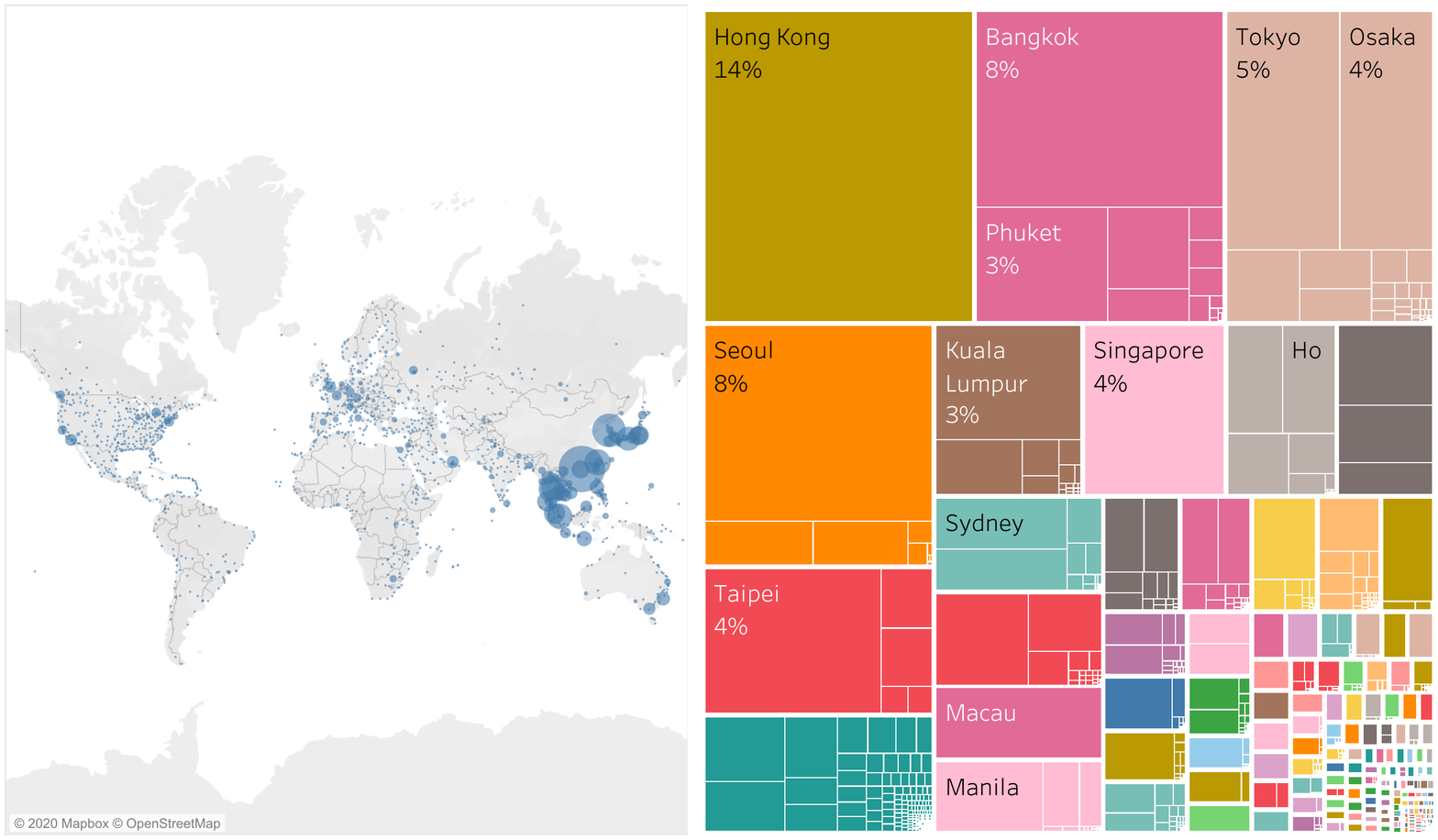}}
\caption{Main destinations outside China for flights originating from China in the period January - April 2020 (baseline without effects from the Coronavirus epidemic).}
\label{fig5}
\end{figure}

Outside China the main destinations are the Special Administrative Region of Honk Kong (14\%), Thailand (13\%), Japan (11\%) and South Korea (9\%). Within these countries the main airports are Hong Kong (14\%), Bangkok (8\%), Seoul (8\%), Tokyo (5\%) and Taipei (4\%) (Figure~\ref{fig5}).

\begin{figure}[!htb]
\centering{\includegraphics[width=1\textwidth,height=0.4\textheight]{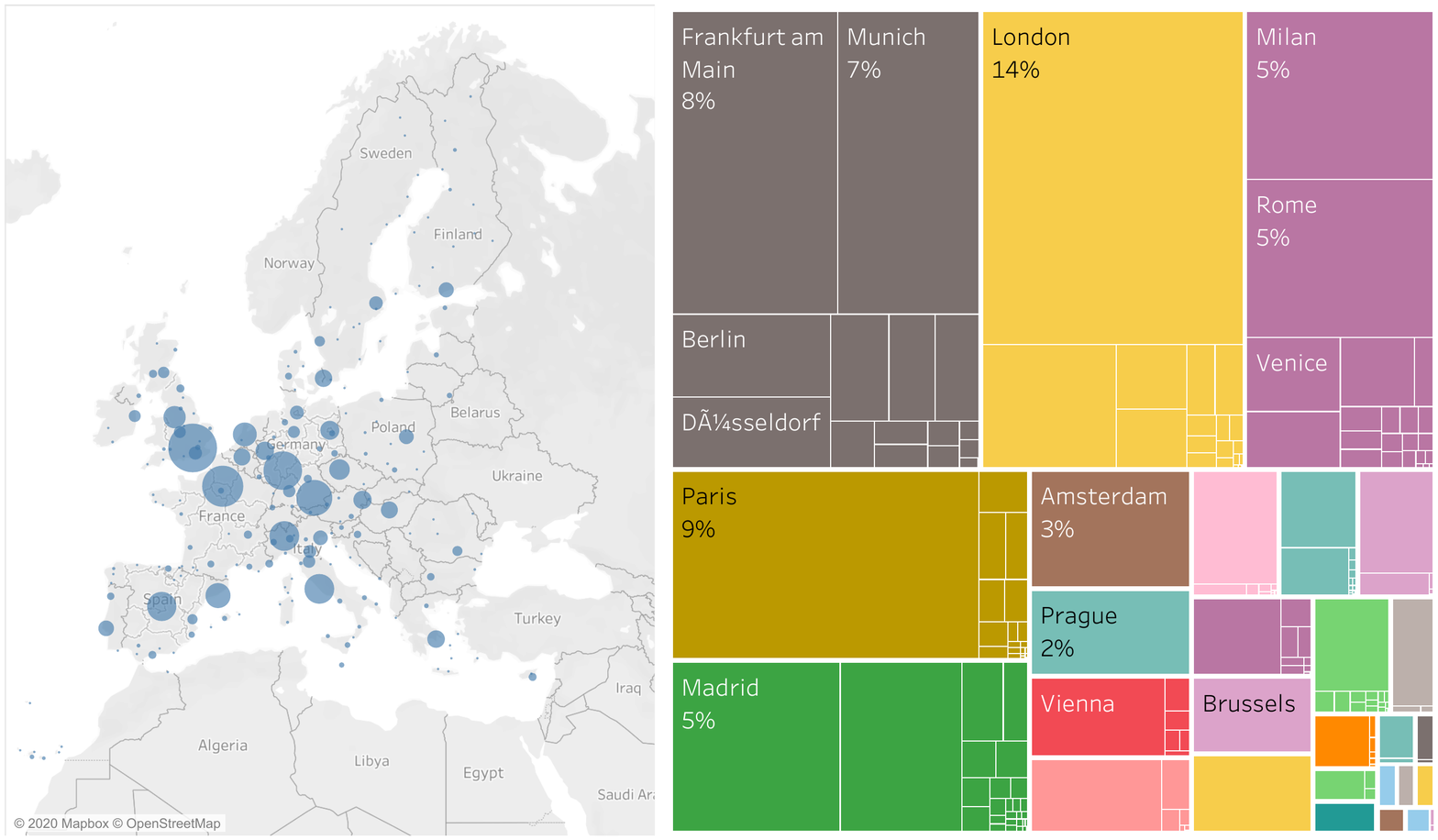}}
\caption{Main destinations in EU countries and UK for flights originating from China in the period January - March 2020 (baseline without effects from the Coronavirus epidemic).}
\label{fig6}
\end{figure}

In the case of the EU and UK the largest flows are directed to airports in Germany (Frankfurt, 8.1\%), UK (London, 14.1\%), Italy (Milan, 5.1\%) and France (Paris, 9.4\%) (Figure~\ref{fig6}). 

\begin{figure}[!htb]
\centering{\includegraphics[width=1\textwidth,height=0.4\textheight]{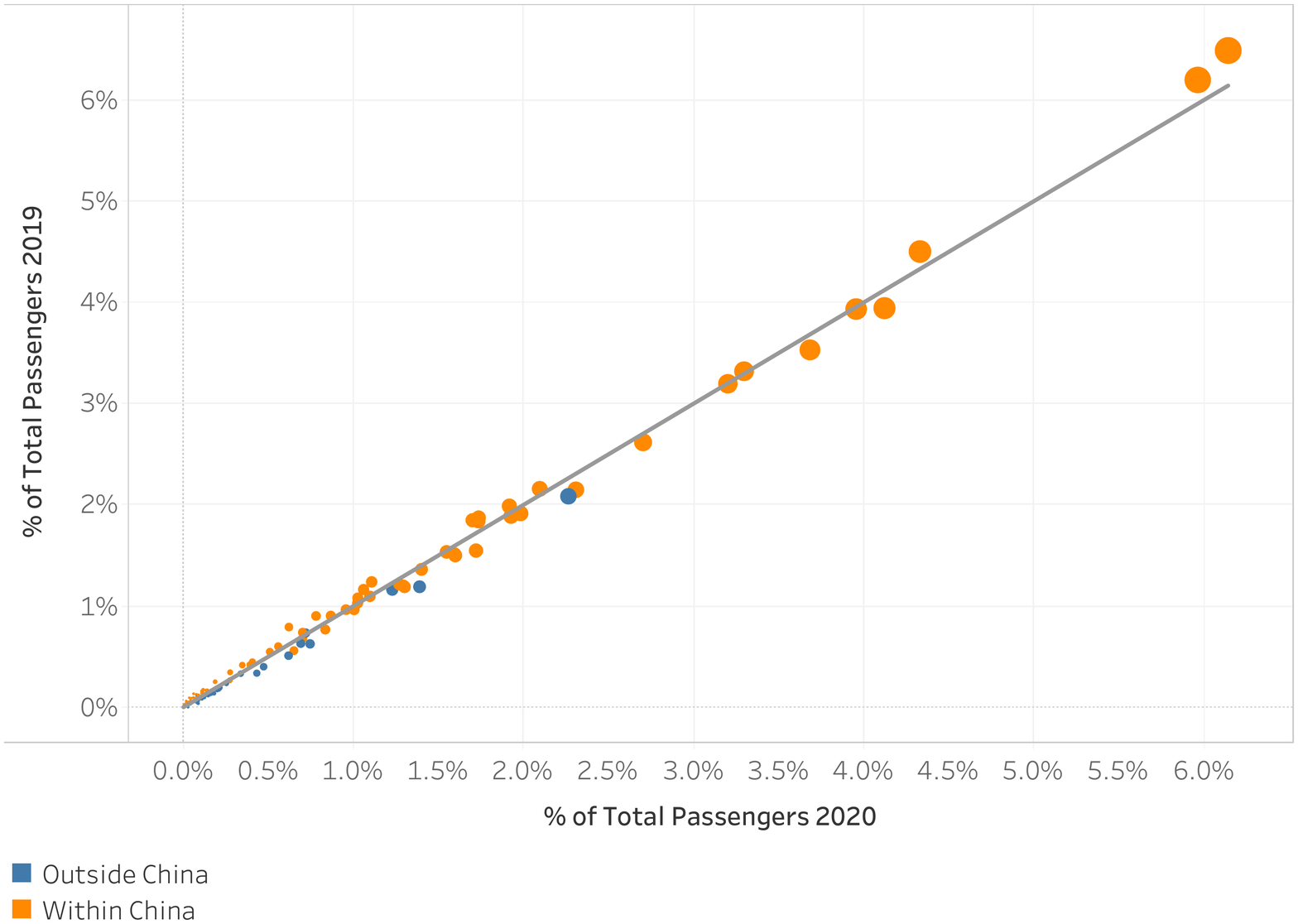}}
\caption{Comparison of the share of flows for each airport of destination between January-March 2019 and January - March 2020. Alignment along the diagonal indicates that the baseline estimates for 2020 are reproducing the distribution of flows by airports in the historical data.}
\label{fig7}
\end{figure}

Figure \ref{fig7} shows the share of flows for each airport of destination in the period January-March 2019 in comparison to the period January - March 2020. 
The alignment along the diagonal indicates that our estimates for the period January - March 2020 largely reflect the geographical distribution in 2019.

\section{Reduction in air passengers flows following the complete closure of routes}
\begin{figure}[!htb]
\centering{\includegraphics[width=1\textwidth,height=0.4\textheight]{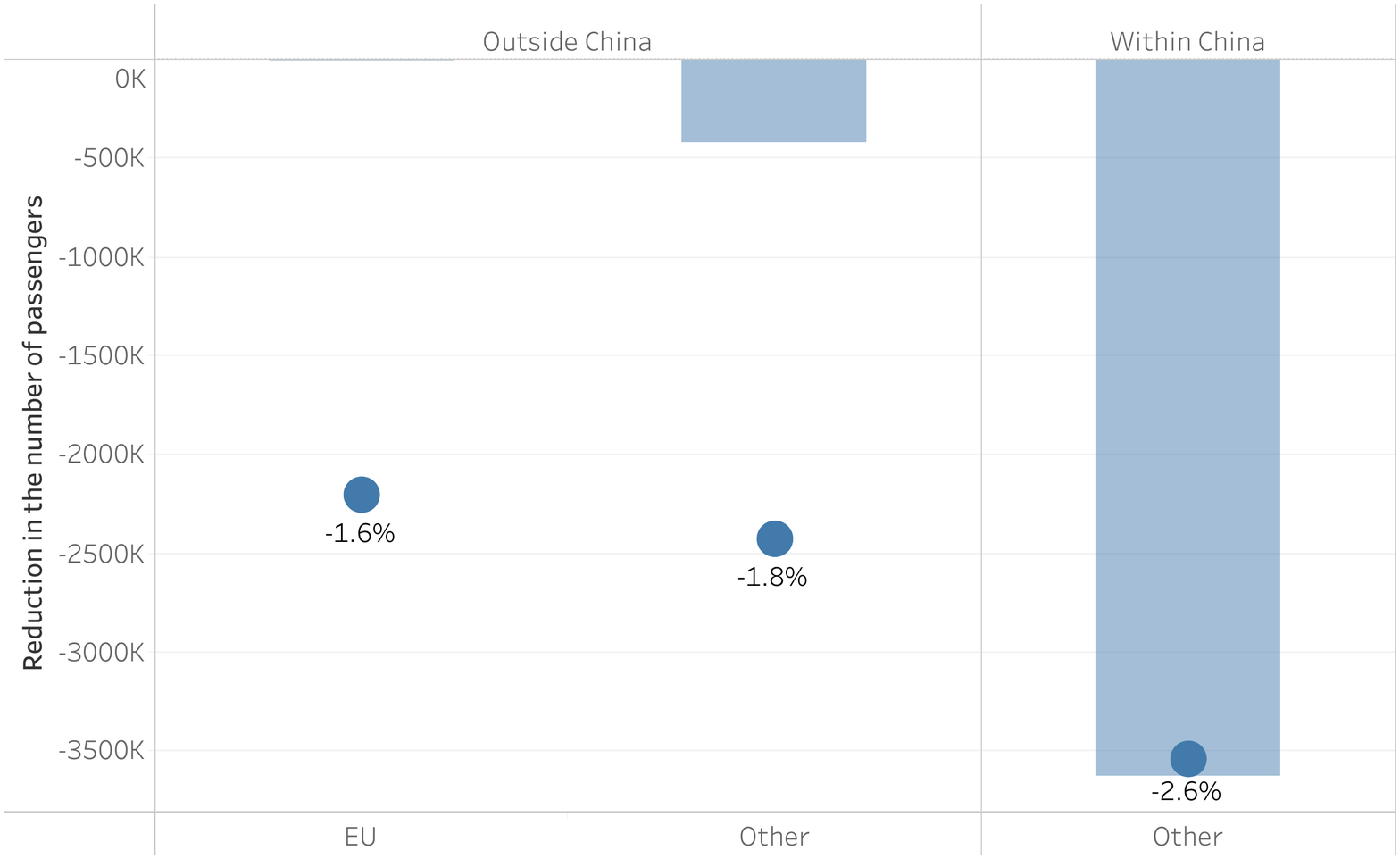}}
\caption{Absolute and relative reduction in the number of passengers, following closure of routes during the Coronavirus epidemic in the period January - March 2020.}
\label{fig8}
\end{figure}

After having established a baseline for the period October 2019 - March 2020 our main goal was to get first signals of the reduction in the number of passengers after the Coronavirus outbreak. As indicated in the section on methods, we estimate the reduction of passengers only in cases for which the online booking systems do not provide any option of connection.
Overall the reduction for all flights originating from China in the period January - March 2020 is estimated at -4 million passengers, corresponding to -2.5\% of the expected volume of traffic.
Figure~\ref{fig8} shows that the reduction in the number of passengers is mostly taking place within China (-2.6\%).

\begin{figure}[!htb]
\centering{\includegraphics[width=1\textwidth,height=0.4\textheight]{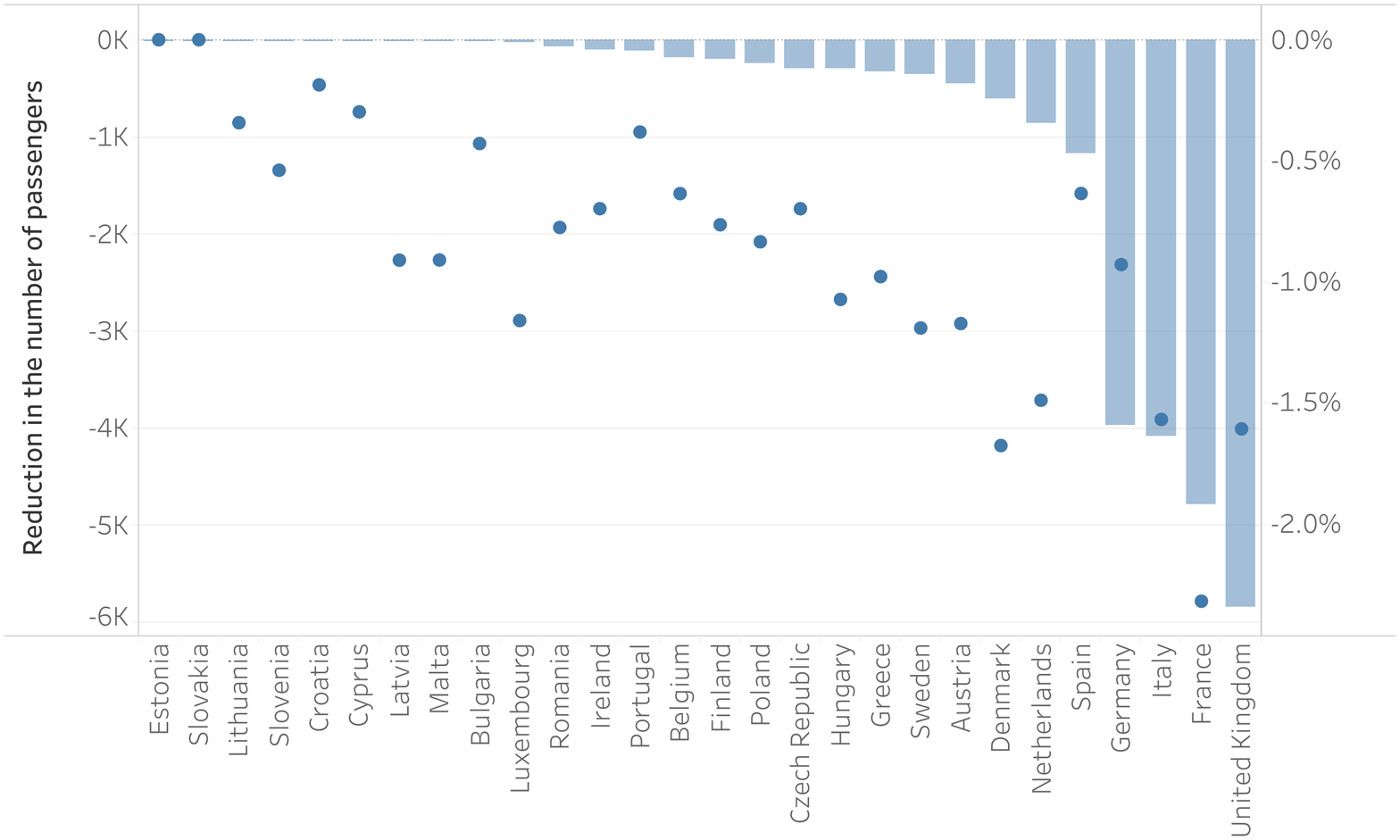}}
\caption{Absolute and relative reduction in the number of passengers towards EU destinations and UK, following closure of routes during the Coronavirus epidemic in the period January - March 2020.}
\label{fig9}
\end{figure}

Figure~\ref{fig9} provides further details for destination in the EU and UK. In absolute terms the highest reduction is for the UK with -6 thousand passengers (-1.6\%), followed by France (-2.3\%), Italy (-1.6\%) and Germany (-0.9\%).

\pagebreak

\section*{Figures}

\bibliography{ref}

\bibliographystyle{chicago}

\end{document}